\documentclass[10pt,prl,aps,twocolumn,superscriptaddress,floatfix]{revtex4-1}
\usepackage{amsmath,amssymb} 
\usepackage{txfonts,graphicx,bm,color}

\begin{document}

\title{Magnetic Structures and Spin-wave Excitations in Rare-Earth Iron Garnets near the Compensation Temperature}
	
\author{Michiyasu Mori}
\affiliation{Advanced Science Research Center, Japan Atomic Energy Agency, 
	Tokai, Ibaraki 117-1195, Japan} 

\author{Timothy Ziman}
\affiliation{Institut Laue Langevin, 38042 Grenoble Cedex 9, France}
\affiliation{Institute for Materials Research, Tohoku University, Sendai, Miyagi 980-8577, Japan}

\begin{abstract}
We introduce a simple model for the ferrimagnetic non-collinear ``magnetic umbrella" states of  rare-earth iron garnets (REIG), common when the  rare-earth moments have non-zero orbital angular momentum. The spin-wave excitations  are calculated   within linear spin wave theory and temperature effects via mean-field theory.  This could be used to determine the magnetic polarization of each mode and thereby the spin currents generated by thermal excitations including the effects of mixed chirality.  The spectra  reproduce essential features seen in more complete models, with hybridization between the rare earth crystal field excitations and the propagating mode on the iron moments. By the symmetry of the model, only one rare earth mode hybridizes, inducing a gap at zero wave number and level repulsion at finite frequency. At the compensation point, the hybridization gap closes and finally, as we approach the N\'eel temperature, the hybridization gap appears to reopen. 
The chirality of the lowest mode changes its sign around the frequency at which the level repulsion occurs. This is important to estimate the spin current generation in REIGs.
\end{abstract}

\date{2023.3.27}

\maketitle

\section{Introduction}
A {\it ferri}magnet as a kind of {\it ferro}magnet composed of sublattices of  anti-parallel, but unequal moments,  was  predicted  theoretically by N\'eel\cite{neel48}. 
Soon afterwards, the phenomenon of magnetic compensation,  where
the magnetization vanishes at a compensation temperature $T_{\rm c}$ far below the N\'eel temperature $T_{\rm N}$~\cite{gorter53} was observed in the spinel ferrite LiFeCr. 
Such a ferrimagnet, called an N-type ferrimagnet,  has also been found in rare-earth iron garnets (REIGs)\cite{neel63,pauthenet54}. 
The REIGs have been studied by many authors,  with the aim of applying their magnetization-compensation properties to magneto-optical memories\cite{chang65,chow68,nelson68}. 
In addition to the magnetization-compensation, an angular momentum compensation at $T_A\ne T_c<T_N$ is observed using the Barnett effect~\cite{imai19,chudo21}. Near $T_A$, it is reported that the magnetizations reverse rapidly and the domain walls move fast\cite{stanciu07,kim17}. Those properties are advantageous for magnetic memories.
 
Spin-wave excitations in yttrium iron garnet, where only the iron atoms bear moments, were studied in detail by Nambu et al. using inelastic neutron scattering\cite{nambu20}.
They found two modes, acoustic and optical, having opposite spin polarization (chirality).
Notably, the optical gap is gradually suppressed by increasing temperature $T$~\cite{plant77}. 
Furthermore in the case of N-type REIG, e.g., gadolinium iron garnet, the optical mode is calculated to merge both to the acoustic mode and to the crystal field excitations near $T_c$~\cite{geprags16}. 
This strongly affects the spin current generation, since it is determined by the magnon populations of the those modes. 
On the other hand, it is known that most  of the REIG, where the rare-earth ion has an orbital contribution to the total moment and therefore strong crystal field effects, have a non-collinear magnetic structure called an ``umbrella state"\cite{pickart70}. 
In the magnetic umbrella state, the chirality of spin-wave excitations will be mixed by the hybridization of  each mode, which now has amplitudes on different ions  that are ordered in non-collinear directions. As the spin wave excitation is locally transverse to the ordered moment, exchange terms mix the chirality as they propagate from site to site\cite{tomasello22}. 
If this is the case,  the magnitude of the generated spin current cannot be a simple sum of the magnon populations over different bands. 
To clarify the details of the spectral weight of spinwave excitations in the REIG, we propose a simple model to reproduce several of the essential features of the magnetic structure of REIG. 
Our model shows a simplified form of an umbrella state that is planar in spin space. 

\section{Model Hamiltonian and Mean-field Solution}
\begin{figure}[h]
	\centering
	\includegraphics[width=0.4\textwidth]{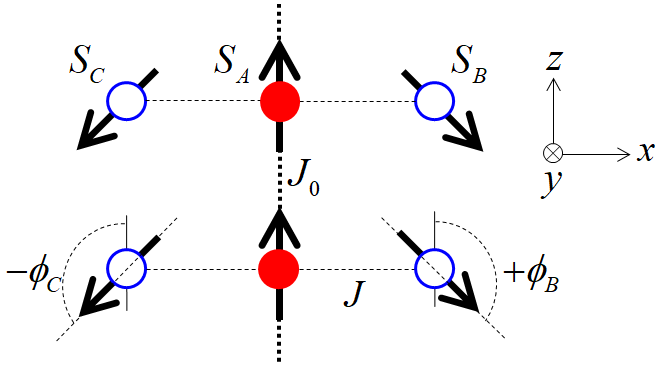}
	\caption{Lattice structure of the model reproducing the planar version of the umbrella state. 
	Both angles $\phi_B$ and $\phi_C$ are defined in the range of $0\le \phi_B, \phi_C \le \pi$.}
	\label{lattice-l}
\end{figure}

\begin{figure}[h]
	\includegraphics[width=0.45\textwidth]{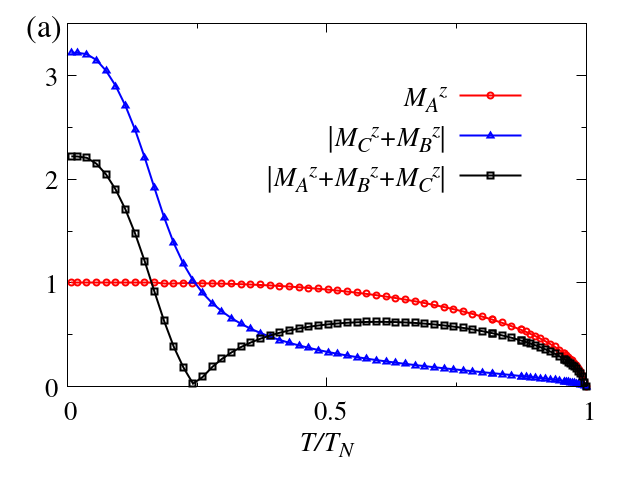}
	\includegraphics[width=0.45\textwidth]{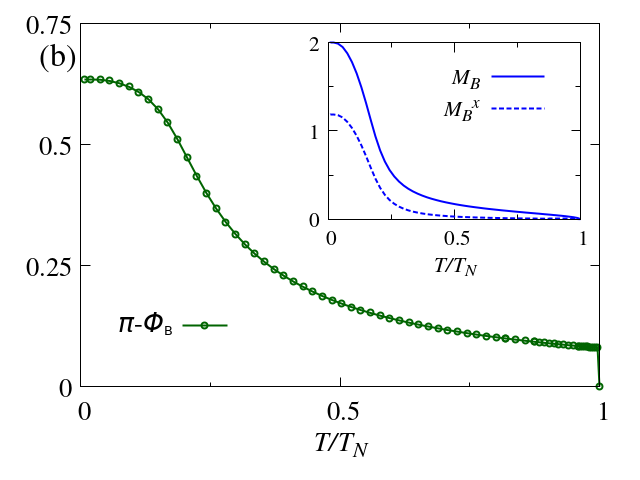}
	\caption{The mean-field solution for $j=0.05$, $d=0.025$, 
		$|S_A|=1$, and $|S_B|=|S_C|=2$.  In (a) the net magnetization in the axes of the Iron moments (blue) as well as the contributions from the Fe (in red)  and rare-earth ( in blue)  (b) The canting angle of the rare-earth ions away from the direction antiparallel to the net iron moment. While it decreases towards the compensation temperature, it actually remains finite at all temperatures below the N\'eel temperature. In inset, the total  (solid line) and transverse (dashed)  components of the thermally averaged magnetization on the rare-earth ions. }
	\label{mfsol}
\end{figure}

The physical REIG in their cubic phase have 64 magnetic ions in a primitive cell with a ratio of three rare-earth ions RE$^{+3}$ to five Fe$^{3+}$. 
For  the five Fe$^{3+}$ sites, three are on tetrahedral $d$-sites and  two  on octahedral $a$-sites. Recently Tomasello {\it et al.}\cite{tomasello22} introduced a simple model respecting stoichiometry with eight sites to understand the essence of the magnetic umbrella structure and dynamics within a spin-wave approximation.  The reduction from the original structure meant that in that model one has only to solve 16  by 16 matrices, which can be done with rather light numerics.  Since the  antiferromagnetic  couplings between the neighbouring $d$ and $a$ sites are an order of magnitude stronger than any other, for some purposes and  especially for temperatures well below $T_N$, we can attempt to reduce  the model even further by associating the cluster of five Fe$^{3+}$ sites  with a single  effective magnetic moment, which is what we implement in this paper.
The magnetic moment reduced from five Fe$^{3+}$ sites   to a single spin couples antiferromagnetically to the R$^{+3}$ sites, which are subject to an anisotropy that tilts  from the quantization axis of Fe$^{3+}$ sites.  To make the model less realistic, but even more tractable,
we propose here a  minimal model by reducing the three ``ribs" of the umbrella  to two, giving a two-dimensional ``flatland umbrella" instead of the three-dimensional spin structure. This enables us to develop intuition into  the general  effects of crystal field induced non-collinearity with  even simpler  algebra than previously. It is hard to see how to reduce such a model further: the two ``ribs" with orthogonal easy axes for each effective iron moment can balance out to give a non-collinear structure even with isotropic exchange.
Note that a model with two sets of spins with orthogonal axes was
	originally proposed by Moriya to describe the dynamics
	of NiFe$_2$~\cite{moriya60}. Thus the model here can be considered a generalization of Moriya's, with  the two sets coupled to an extra set of spins without anisotropy.
The Hamiltonian is given by 
\begin{align}
	H &=  
	- J_0\sum_{\langle i,j \rangle}\vec S_{A,i} \cdot \vec S_{A,j}
	+ J \sum_{i,p =B,C} \vec S_{A,i} \cdot \vec S_{p,i} \nonumber\\
	&- D\sum\limits_i {{{\left( { - S_{B,i}^x + S_{B,i}^z} \right)}^2}}  - 
	D\sum\limits_i {{{\left( {S_{C,i}^x + S_{C,i}^z} \right)}^2}},\label{heisenberg}
\end{align}
with the spin operators $\vec{S}_{p,i}$ on $i$-th site of $p$-site 
($p=A,B,C $) as shown in Fig. \ref{lattice-l}. 
The angular brackets $\langle\cdots\rangle$ denotes nearest neighbor sites. 
The magnitudes of the magnetic exchange interactions $J_0$, $J$, and the 
anisotropic energy $D$ are positive. 
Both angles $\phi_B$ and $\phi_C$ are defined in the range of $0\le \phi_B, \phi_C \le \pi$.
The mean field Hamiltonian per site leads to,
\begin{align}
	h_{\rm MF}
	&=  - 2{{\vec M}_A} \cdot {{\vec S}_A}
	+ j \sum\limits_{p = B,C} {\left( 
		{{{\vec M}_A} \cdot {{\vec S}_p} + {{\vec M}_p} \cdot {{\vec S}_A}} 
		\right)} \nonumber\\
	&- 2d\left( { - M_B^x + M_B^z} \right)\left( { - S_B^x + S_B^z} \right)\nonumber\\
	&- 2d\left( {M_C^x + M_C^z} \right)\left( {S_C^x + S_C^z} \right),\label{hmf}
\end{align}
where $j\equiv J/J_0$, and $d\equiv D/J_0$. 
The energy is scaled by $J_0$. 
The expectation values of spins, $\vec M_p=\langle \vec S_p \rangle$, 
are self-consistently calculated using the Hamiltonian (\ref{hmf}).
The mean-field solutions for $j=0.05$, $d=0.025$, and  
$|S_A|=1$, $|S_B|=|S_C|=2$, are plotted as a function of $T$ in Fig. \ref{mfsol}.
The non-collinear magnetic structure continues up to the $T_N$. 
The magnetization compensates around $T_c/T_N\sim 0.25$ as shown in Fig.~\ref{mfsol} (a). 
Simultaneously, the angles $\phi_B$ ($=\phi_C$) (Fig.~\ref{mfsol} (b)) start to 
decrease rapidly around $T_c$. 
Although the magnetization on the $B$ and $C$ sites rapidly decreases, their angles are still finite up to $T_N$ within the mean field approximation. However, Lahubi {\it et al.} reported that the satellite peaks relevant to the umbrella structure were suppressed at $T_c$~\cite{lahoubi84}. 
The fluctuations may suppress the umbrella structure.

\section{Spinwave Excitations}
\begin{figure}[htb]
	\includegraphics[width=0.4\textwidth]{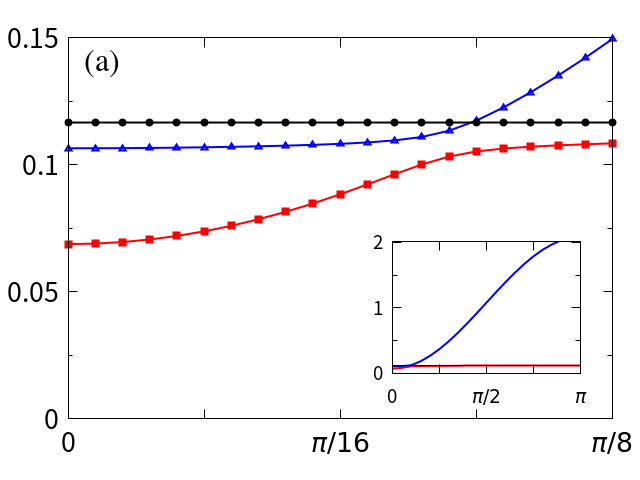}
	\includegraphics[width=0.4\textwidth]{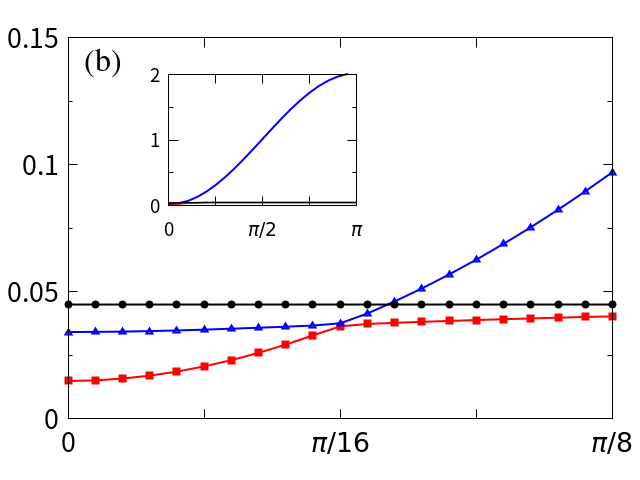}
	\includegraphics[width=0.4\textwidth]{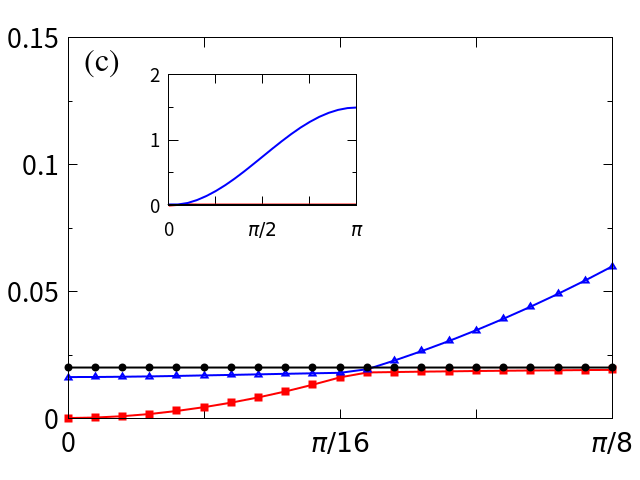}
	\caption{The dispersion relation of spinwave excitations: at low temperatures (a) $T/T_N\sim 0$,  at the compensation temperature (b) $T\sim T_c$, and (c)  approaching the N\'eel temperature $T/T_N\sim 0.75$. Each inset is the dispersion relation over the full reciprocal space. }
	\label{disp}
\end{figure}

What will happen in spinwave excitations?
The Holstein-Primakoff (HP) bosons (magnons) describe fluctuations around the mean field solution. 
We need to take care of the expansions on the $B$- and $C$-sites, since their magnetizations are not in the global $z$-direction.
To treat the spinwaves in the non-collinear magnetic order, 
the quantization axes on $B$- and $C$-lattices are rotated to make their magnetization aligned in the local $z$-direction by, 
$\vec S_B'=R_y(\phi_B)\vec S_B$ and $\vec S_C'=\hat R_y(-\phi_C)\vec S_C$. 
The rotation matrix around the $y$-axis is denoted by $\hat R_y(\phi_p)$ ($p=B, C$). 
Rotating the axes, the Hamiltonian (\ref{heisenberg}) scaled by $J_0$ is given by, 
\begin{align}
	h &=  
	- \sum_{\langle i,j \rangle}\vec S_{A,i} \cdot \vec S_{A,j}\nonumber\\
	&+ j \sum_{i} \vec S_{A,i} \hat R_y^{-1}(\phi_B) \cdot \vec S_{B,i}' 
	+ j \sum_{i} \vec S_{A,i} \hat R_y^{-1}(-\phi_C) \cdot \vec S_{C,i}' \nonumber\\
	&- d\sum_i \left[-(\cos\phi_B+\sin\phi_B)S_{B,i}'^x+ (\cos\phi_B-\sin\phi_B)S_{B,i}'^z \right]^2\nonumber\\
	&- d\sum_i \left[(\cos\phi_C+\sin\phi_C) S_{C,i}'^x + (\cos\phi_C-\sin\phi_C)S_{C,i}'^z \right]^2.
	\label{rotated}
\end{align}
The Holstein-Primakoff bosons--for which the creation and annihilation operators are $a_i^\dag, a_i$ on $A$-sublattice,  
$b_j^\dagger, b_j$ on $B$-sublattice, and 
$c_i^\dag, c_i$ on $C$-sublattice--are given by,
$S_{A,i}^-=\sqrt{2M_A} a_i^\dag$, 
$S_{A,i}^+=\sqrt {2M_A}a_i$, 
$S_{A,i}^z=M_A-a_i^\dag {a_i}$, 
$S_{B,i}'^+=\sqrt{2M_B}b_i$, 
$S_{B,i}'^-=\sqrt{2M_B}b_i^\dag$, 
$S_{B,i}'^z=M_B-b_j^\dag {b_i}$, 
$S_{C,i}'^+=\sqrt{2M_C}c_i$, 
$S_{C,i}'^-=\sqrt{2M_C}c_i^\dag$, 
$S_{C,i}'^z=M_C-c_j^\dag {c_i}$, 
with $M_p\equiv \sqrt{(M_p^x)^2+(M_p^z)^2}$. 
The spinwave Hamiltonian in the momentum space $h_{SW}$ is given by,
\begin{align}
	h_{SW}
	&=
	\sum\limits_k {{\varepsilon _A}a_k^\dag {a_k}}  
	+\sum\limits_k {{\varepsilon _B}} b_k^\dag {b_k} 
	+\sum\limits_k {{\varepsilon _C}} c_k^\dag {c_k}\nonumber\\
	&+ \sum\limits_k {\left[ {\tau_{AB}\left( {{a_k}{b_{ - k}} + a_k^\dag b_{ - k}^\dag } \right) + {\eta _{AB}}\left( {a_k^\dag {b_k} + {a_k}b_k^\dag } \right)} \right]} \nonumber\\
	&+ \sum\limits_k {\left[ {{\tau _{AC}}\left( {{a_k}{c_{ - k}} + a_k^\dag c_{ - k}^\dag } \right) + {\eta _{AC}}\left( {a_k^\dag {c_k} + {a_k}c_k^\dag } \right)} \right]} \nonumber\\
	&+ \sum\limits_k {{\delta _B}\left( {{b_k}{b_{ - k}} + b_k^\dag b_{ - k}^\dag } \right)}  + \sum\limits_k {{\delta _C}\left( {{c_k}{c_{ - k}} + c_k^\dag c_{ - k}^\dag } \right)},\label{sw}
\end{align}
where $k$ denotes the momentum. 
In the above equations, 
\begin{align}
	{\varepsilon _A} 
	&\equiv 2{M_A}\left( {1 - \cos k} \right) - j\left(M_B \cos \phi_B+ M_C \cos \phi_C \right),\\
	{\varepsilon _B} 
	&\equiv  
	- j{M_A}\cos {\phi _B}
	+ dM_B(1 - 3\sin 2{\phi _B}),\\
	{\varepsilon _C} 
	&\equiv 
	- j{M_A}\cos {\phi _C}
	+ dM_C(1 - 3\sin 2{\phi _C}),\\
	{\eta _{AB}} &\equiv j{\cos ^2}\frac{{{\phi _B}}}{2}\sqrt {{M_A}{M_B}}, \\
	{\eta _{AC}} &\equiv j{\cos ^2}\frac{{{\phi _C}}}{2}\sqrt {{M_A}{M_C}}, \\
	{\tau _{AB}} &\equiv  - j{\sin ^2}\frac{{{\phi _B}}}{2}\sqrt {{M_A}{M_B}},\\
	{\tau _{AC}} &\equiv  - j{\sin ^2}\frac{{{\phi _C}}}{2}\sqrt {{M_A}{M_C}}, \\
	{\delta _B} &\equiv  - d\frac{{{M_B}}}{2}(1 + \sin 2{\phi _B}),\\
	{\delta _C} &\equiv  - d\frac{{{M_C}}}{2}(1 + \sin 2{\phi _C}).
\end{align}
Due to the mirror symmetry with respect to $A$-sites, 
$M_B=M_C$, 
$\phi_B=\phi_C\equiv \phi$, 
$\varepsilon_B=\varepsilon_C$, 
$\eta_{AB}=\eta_{AC}\equiv \eta$, 
$\tau_{AB}=\tau_{AC}\equiv \tau$,
$\delta_{B}=\delta_{C}\equiv \delta$. 
The constant term is given by,
\begin{align}
	F(\phi) &=  - M_A^2 + 2j M_A M_B \cos\phi -2 dM_B^2(\cos\phi - \sin \phi)^2,
\end{align}
The classical solution is obtained by minimizing $F(\phi)$, 
\begin{align}
	\frac{\partial F(\phi)}{\partial \phi}
	&=-j M_A\sin\phi + 2 d M_B \cos 2\phi =0,\label{min}
\end{align}
which determines the non-collinear structure with $\phi\ne 0$. 
For $d=0$ and $j\ne 0$, the collinear state with $\phi=0$ or $\pi$ is stable. For $j=0$ and $d\ne 0$, the non-collinear state with $\phi=\pi/4$ or $3\pi/4$ is stable.  
Only when Eq.~(\ref{min}) is satisfied, the spinwave approximation is justified. 
Thanks to the symmetry, Eq. (\ref{sw}) is further reduced to, 
\begin{align}
	h_{SW}&=h_+ + h_-,\label{swred}\\
	h_+&= \sum\limits_k {\Psi _k^\dag } 
	\left( 
		\begin{array}{*{20}{c}}
			{{\varepsilon _A}}&{\sqrt 2 \eta}&0&{\sqrt 2 \tau}\\
			{\sqrt 2 \eta}&{{\varepsilon _B}}&{\sqrt 2 \tau}&\delta\\
			0&{\sqrt 2 \tau}&{{\varepsilon _A}}&{\sqrt 2 \eta}\\
			{\sqrt 2 \tau}&\delta&{\sqrt 2 \eta}&{{\varepsilon _B}}
		\end{array}
	\right)	\Psi_k,\label{hplus}\\
	h_-&= \sum\limits_k {\psi _k^\dag } 
	\left( 
		\begin{array}{*{20}{c}}
			{{\varepsilon _B}}&{\delta}\\
			{\delta}&{{\varepsilon _B}}
		\end{array}
	\right) \psi _k,\label{hminus}
\end{align}
where 
$\Psi _k^\dag = \left(a_k^\dag~ f_k^\dag~ a_{-k}~ f_{-k} \right)$,
$\psi _k^\dag = \left( g_k^\dag~ g_{ - k} \right)$, 
$f_k = \left(b_k+c_k\right)/\sqrt{2}$, and
$g_k = \left(b_k-c_k\right)/\sqrt{2}$.
Solving Eq. (\ref{swred}), we found the spinwave excitations in the umbrella state as shown in Figs.~\ref{disp} for (a) $T/T_N\sim 0$, (b) $T_c/T_N\sim 0.25$, and (c) $T/T_N\sim 0.75$.
The blue and red lines are obtained by Eq. (\ref{hplus}), whereas the black line is the non-bonding band by Eq. (\ref{hminus}). The optical mode observed in the experiments does not appear in this calculation, since the anti-parallel iron sites, i.e., $a$- and $d$-sites, are combined into one $A$-site in our model. On the other hand, in Fig.~\ref{disp} (a), we can find the anti-crossing between the acoustic mode (red) and the optical mode (blue), which originates from RE sites, i.e., $B$ and $C$ sites. Near $T_c$, the anti-crossing disappear as shown in Fig.~\ref{disp} (b). Above $T_c$ and near $T_N$, the anisotropy energy becomes irrelevant and the gap in the acoustic mode is suppressed as shown in Fig.~\ref{disp} (c).

\section{Chirality}
\begin{figure}[htb]
	\includegraphics[width=0.4\textwidth]{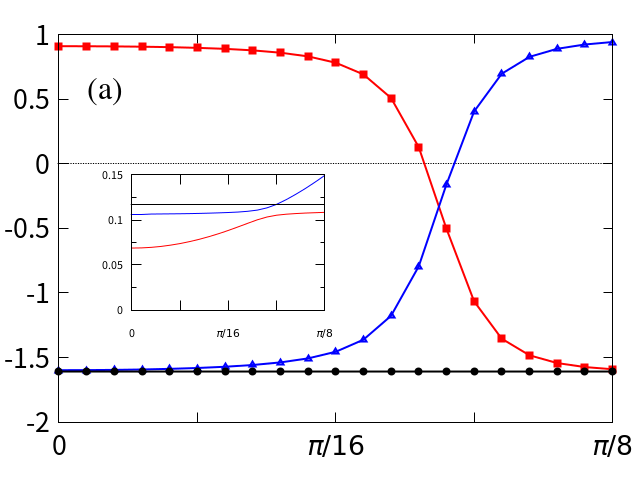}
	\includegraphics[width=0.4\textwidth]{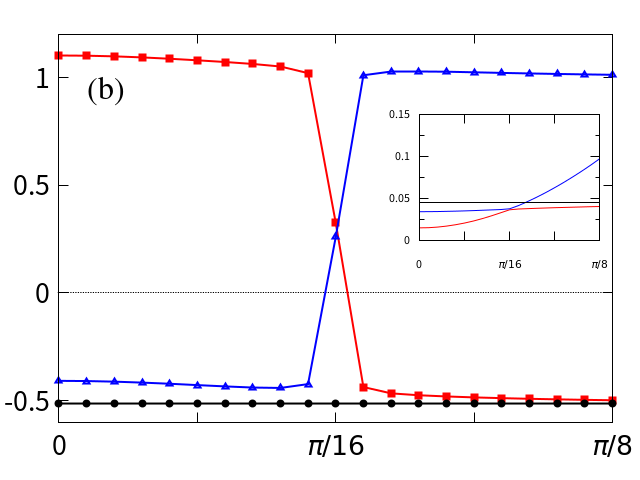}
	\caption{The chirality at low temperatures (a) $T/T_N\sim 0$, and (b) $T\sim T_c$. The red (blue) line is the chirality of the lowest (second) band colored by red (blue) in the inset, which are the dispersion relations shown in Figs. \ref{disp} (a) and (b), respectively.}
	\label{chiral}
\end{figure}

The chirality of each band is estimated using the imaginary part of $\chi_k(\omega)$ with momentum $k$ and frequency $\omega$ given by, 
\begin{align} 
\chi_k(\omega)
	&=\!\frac{1}{2}\!\!\!\sum_{{\tiny l=A,B,C}}\!\!\!\!\!
	 \langle S_{l,k}^+\left(\omega\right)S_{l,-k}^-\left(-\omega\right) 
	 - S_{l,-k}^-\left(-\omega\right)S_{l,k}^+\left(\omega\right)\rangle, \\
	&=M_A\left[g_{aa}(k,\omega)-g_{aa}^\dagger(k,\omega)\right]\nonumber\\
	&+M_B\cos\phi_B\left[g_{ff}(k,\omega)-g_{ff}^\dagger(k,\omega)\right]\nonumber\\
	&+M_B\cos\phi_B\left[g_{gg}(k,\omega)-g_{gg}^\dagger(k,\omega)\right],\\
g_{aa}(k,\omega)
	&\equiv \langle a_k(\omega)a_{-k}^\dagger(-\omega)\rangle,\label{ga}\\
g_{ff}(k,\omega)
	&\equiv \langle f_k(\omega)f_{-k}^\dagger(-\omega)\rangle,\label{gf}\\
g_{gg}(k,\omega)
	&\equiv \langle g_k(\omega)g_{-k}^\dagger(-\omega)\rangle\label{gg}
\end{align}
in which $M_B=M_C$ and $\phi_B=\phi_C$ are imposed. 
The thermal Green's functions of HP-bosons $g_a(k,i\omega_n)$, $g_f(k,i\omega_n)$, and $g_g(k,i\omega_n)$ are obtained using Eqs.(\ref{hplus}) and (\ref{hminus}) as, 
\begin{align}
&\left( 
\begin{array}{*{20}{c}}
	g_{aa}(k,i\omega_n)&g_{af}(k,i\omega_n)&0&0\\
	g_{fa}(k,i\omega_n)&g_{ff}(k,i\omega_n)&0&0\\
	0&0&g_{aa}^\dag(k,i\omega_n)&g_{af}^\dag(k,i\omega_n)\\
	0&0&g_{fa}^\dag(k,i\omega_n)&g_{ff}^\dag(k,i\omega_n)
\end{array}
\right)\nonumber\\
	=& 
	\left( 
	\begin{array}{*{20}{c}}
		{-i\omega_n+{\varepsilon _A}}&{\sqrt 2 \eta}&0&{\sqrt 2 \tau}\\
		{\sqrt 2 \eta}&{-i\omega_n+{\varepsilon _B}}&{\sqrt 2 \tau}&\delta\\
		0&{-\sqrt 2 \tau}&{-i\omega_n-{\varepsilon _A}}&{-\sqrt 2 \eta}\\
		{-\sqrt 2 \tau}&-\delta&{-\sqrt 2 \eta}&{-i\omega_n-{\varepsilon _B}}
	\end{array}
	\right)^{-1}&\label{gplus}\\
&\left( 
\begin{array}{*{20}{c}}
	g_{gg}(k,i\omega_n)&0\\
	0&g_{gg}^\dag(k,i\omega_n)
\end{array}
\right)	\nonumber\\
	&= 
	\left( 
	\begin{array}{*{20}{c}}
		-i\omega_n+\varepsilon _B&{\delta}\\
		{-\delta}&{-i\omega_n-{\varepsilon _B}}
	\end{array}
	\right)^{-1},\label{gminus}
\end{align}
where $\omega_n$ is the Matsubara frequency of bosons. The imaginary part of Green's function is obtained by the retarded Green's functions using  analytical continuation. 
Note that the chirality corresponds to the spectral weight  of Eqs. (\ref{ga}), (\ref{gf}) and (\ref{gg}).  
The chirality of each band is plotted in Fig. \ref{chiral}  for (a) $T/T_N\sim 0$, and (b) $T\sim T_c$. 
In both cases, the non-bonding band colored by black is always negative. On the other hand, the lowest band, traced in red in the figures, changes its sign from positive to negative and the blue band has the opposite behavior. It should be noted that the chirality  also changes within one band, but in  a more complex manner, in the model studied by Tomasello et al., \cite{tomasello22}. Near the compensation temperature, $T\sim T_c$, the sign of each bands (red and blue)  changes  sharply near the band-crossing momentum. 
The chirality change within one mode is induced by the non-collinear magnetic structure. Since the quantization axes are not parallel each another, two modes with different chirality can be mixed. This does not happen in a collinear ferrimagnet such as yttrium iron garnet. The simpler chirality dependence  we find compared to  the model in Ref. \cite{tomasello22} may reflect the fact that our ordered state, while non-collinear, remains planar.
As chirality mixing is important to estimate the spin current generation in REIGs, it will be important to understand the detailed form in simplified models, and with the complete spin structure.

\section{Summary}
We studied magnetic structures of a ferrimagnet using a simple model reproducing important features of the  rare-earth iron garnets (REIGs), which have a non-collinear magnetic structure called  an ``umbrella state". The spin-wave excitations of REIG in the umbrella state are calculated with the ultimate aim to estimate the magnetic polarization of each mode. 
In our result, the chirality of the lowest mode changes its sign around the frequency  at which  level repulsion occurs.
Both the dispersion and chiralities of the spin-waves are important in order to know the spin current generated in the REIG. While the present model is highly simplified, even more so than that previously proposed \cite{tomasello22}, it has the advantage of being extremely simple to solve, allowing us to develop intuition both for the more complete description in the previous model or the linear spin wave description for the full crystal structure, which should be obtained from heavier numerical procedures. The model is  very easily generalisable, in particular to incorporate higher spatial dimensionality, necessary so that fluctuations can be incorporated. Obviously the mean field approximation used for finite temperatures here cannot be justified in a one-dimensional model, so implicitly our conclusions suppose that interchain couplings are actually present. 
\par
The interesting results of the calculations here can be seen by examination of  Figures 2, 3 and 4: as in  the model of \cite{tomasello22} a  non-collinear structure is predicted to persist at and above the compensation temperature, and in fact even in the limit of vanishing parallel and perpendicular components at $T_N$ the ratio, expressed as a canting angle, has a finite value. In Figure 3 the mirror symmetry of the model gives a completely decoupled flat mode  (in black) as well as two hybridized modes, coming from the original ferromagnetic magnon of the iron moments and the ``crystal field" level on the rare-earth with a gap at zero momentum. 
The hybridization gap opens due to the level repulsion where the unhybridized modes would cross.
Compared to the cubic REIG structure, the two flat dispersion relations replace the multiple degeneracy coming from the 6 inequivalent rare-earth sites in each unit cell. Note that in the model studied in Ref.\cite{tomasello22} there are three inequivalent rare-earth sites, but as in the  model considered  here, only one seems to hybridize strongly with the propagating mode. At the compensation temperature  (Fig. 3b), the gap at zero momentum is reduced and the hybridization gap at a finite momentum closes. 
The  two chiralities of the   anti-crossing levels exchange sign abruptly at this temperature. Finally, on the approach to the N\'eel temperature from below, the gap at zero momentum is further reduced, whereas the hybridization gap appears to re-open with a small magnitude.  
Although the rare earth moments are very small, the angles from the axes of the iron moment are still finite. 
The dynamics in this region would clearly be interesting to study, including fluctuations beyond mean field theory. This model seems like a suitable starting point to make such calculations tractable.

\section*{Acknowledgment}
We would like thank Prof. Fujita, and Drs. Mannix, Gepr\"ags and Tomasello for  valuable discussions. 
This work was supported by JSPS Grant Nos.~JP20K03810, JP21H04987, JP23H***** and the inter-university cooperative research program (No.~202212-CNKXX-0013) of the Center of Neutron Science for Advanced Materials, Institute for Materials Research, Tohoku University.  T.Z. would like to acknowledge support from the Global Institute for Materials Research, Tohoku  University. A part of the computations were performed on supercomputers at the Japan Atomic Energy Agency.

\end{document}